\documentclass[final,3p,times,twocolumn]{elsarticle}

\usepackage{graphicx}

\usepackage{amssymb}
\usepackage{wasysym}

\usepackage{lineno}

\journal{}
\begin{document}
\begin{frontmatter}
\title{Towards energy resolution at the statistical limit from a negative ion time projection chamber}
\author[llnl]{Peter Sorensen}\ead{pfs@llnl.gov}
\author[llnl]{Mike Heffner}\ead{mheffner@llnl.gov}
\author[llnl]{Adam Bernstein}
\author[llnl,lbnl,berk]{Josh Renner}
\author[llnl]{Melinda Sweany}

\address[llnl]{Lawrence Livermore National Laboratory, 7000 East Ave., Livermore, CA 94550, USA}
\address[lbnl]{Lawrence Berkeley National Laboratory, 1 Cyclotron Road, Berkeley, CA 94720, USA}
\address[berk]{Department of Physics, University of California, Berkeley, CA 94720, USA}

\begin{abstract}
We make a proof-of-principle demonstration that improved energy resolution can be obtained in a negative-ion time projection chamber, by individually counting each electron produced by ionizing radiation.
\end{abstract}

\end{frontmatter}

\section{Introduction}
Double beta decay is a rare nuclear decay mode, with a typical mean lifetime $\tau>10^{19}$~y.  It was recently observed in $^{136}$Xe with a half-life $\tau=2\times10^{21}$~y \cite{2011ackerman,2012gando}.  The zero neutrino mode of this decay, $\beta\beta(0\nu)$, appears theoretically possible given that neutrinos have non-zero mass.  Such a decay would violate lepton number conservation and confirm the Majorana nature of the neutrino. It has not been observed, despite significant on-going experimental effort world-wide.  As in any rare-event search, substantial reduction of radioactive backgrounds are a high priority in the search for $\beta\beta(0\nu)$.  Given that the spectral expectation for $\beta\beta(0\nu)$ is a sharp peak at the double beta decay Q-value, excellent energy resolution can be used as an effective discriminant against those backgrounds which remain.  But even in the absence of any radioactive background, the sensitivity of a search for $\beta\beta(0\nu)$ may be limited by the high energy tail of the spectrum from the two neutrino decay mode, via the finite energy resolution of the detector.  The fraction of $\beta\beta(2\nu)$ counts which contaminate the $\beta\beta(0\nu)$ signal region is given by \cite{2002elliott}
\begin{equation}
F=\frac{kQ\delta^6}{m_e}
\end{equation}
where $m_e$ is the electron mass, $Q=2458$~keV is the endpoint energy of the decay, $\delta$ is the FWHM energy resolution at the endpoint and $k$ is a coefficient which depends weakly on $\delta$. As an example, suppose a  xenon detector obtains $\delta=0.05$ (equivalently, a Gaussian $\sigma=2.1\%$).  From \cite{2002elliott}, we find $k=7$. In a 20 tonne-year exposure (typical of a next generation experiment such as LZ20 \cite{2011malling}), 15 $\beta\beta(2\nu)$ background events would be expected in the $\beta\beta(0\nu)$ signal region. Larger values of $\delta$ quickly lead to larger leakage. Only upper limits on the half-life for $\beta\beta(0\nu)$ exist at present. This suggests that there is a significant motivation for developing a new type of xenon detector whose energy resolution approaches the statistical limit.

\begin{figure*}[ht]
\centering
\includegraphics[width=0.95\textwidth]{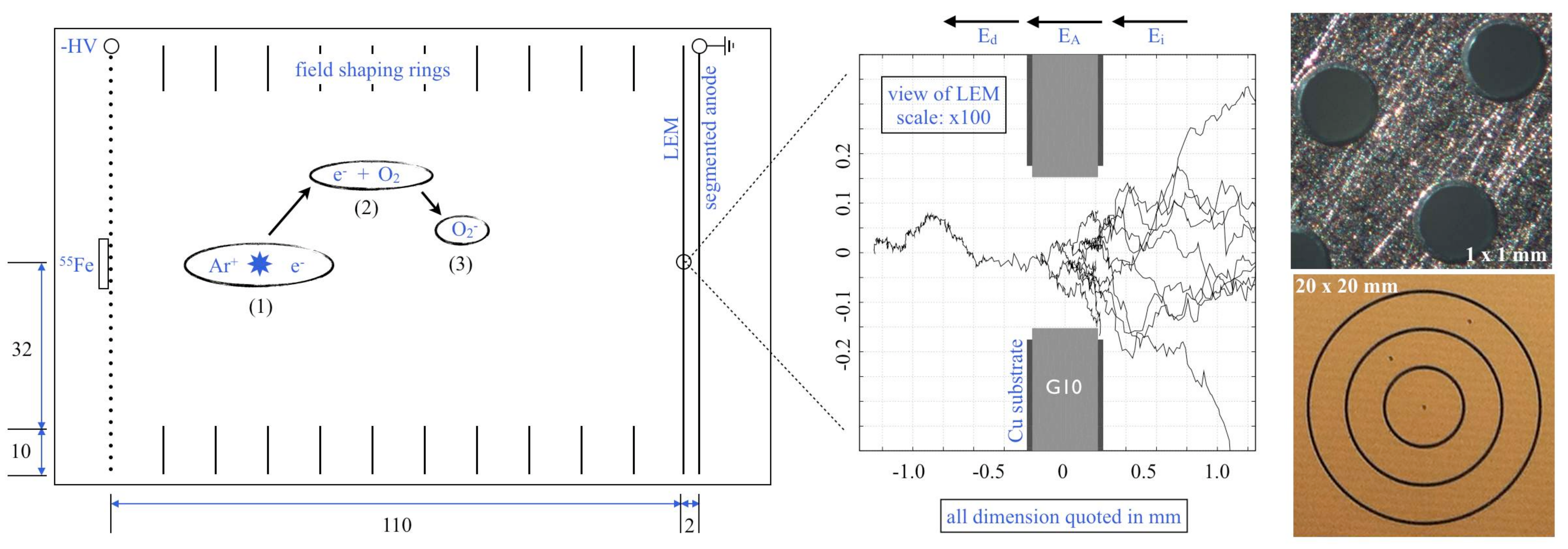}
\caption{{\bf (Left)} Cross-section schematic of the detector, showing (1) initial generation of electron-ion pairs, (2) rapid electron capture on oxygen and (3) negative ion drift.  {\bf (Center)} Enlarged cross-section schematic of the LEM, showing a simulated electron avalanche (with gain $\times30$) as obtained from the Garfield software \cite{garfield}. {\bf (Right)} Close-up photographs of the LEM (top) and segmented anode (bottom).}
\label{fig1}
\end{figure*}

In this work we make a proof-of-principle demonstration that improved energy resolution can be obtained by individually counting each electron produced by ionizing radiation. The idea has been previously suggested \cite{2005martoff,2007nygren} but to our knowledge never implemented.  Our detector is a negative ion time-projection chamber, shown schematically in Fig.~\ref{fig1}.  Rather than drift and detect the electrons created by ionizing radiation, we introduce oxygen as an electronegative dopant to capture the electrons. Oxygen ions O$_2^-$, rather than electrons, are then drifted to the anode.  Just prior to arrival at the anode, the electrons are recovered from the O$_2^-$ and amplified, using a large electron multiplier (LEM) \cite{2009breskin}. Ideally, each electron is then individually counted.  

The dopant concentration and gas density are tuned so that electron capture occurs over a range of $\sim$cm.    We note that previous negative ion time-projection chambers have used either pure carbon disulfide (CS$_2$) \cite{2000martoff}, mixtures of CS$_2$ with noble gases \cite{2000snowden,2001ohnuki} or nitromethane (CH$_3$NO$_2$) \cite{2009martoff} dopant. In this work, we employed a mixture of argon (66\%,  as a low-cost surrogate for xenon), carbon dioxide (30\%, as quench gas \cite{1999bressan}) and oxygen (4\%) as electronegative dopant.

\section{Apparatus} \label{sec:exp}
The construction of our detector is shown schematically in Fig.~\ref{fig1}.  An enlarged view of the region around the LEM is also shown, with a simulated electron avalanche of gain $\times30$.  The LEM was constructed from copper-clad G10 printed circuit board of thickness 0.432~mm with 0.305~mm diameter holes.  After drilling, a chemical etch was used to remove 0.023~mm of copper from the circumference of each hole, in order to increase the breakdown voltage.  A segmented anode consisted of four concentric rings of radius 3, 6, 9 and 40~mm, with 0.25~mm standoff between rings.  Signals were digitized from the center three anodes, and the large outer ring was simply maintained at ground potential.  The rings were used as veto to ensure complete collection of all electrons from each event.  


A charge-sensitive preamplifier with a gain of 1.4~V~pC$^{-1}$ and recovery time constant $\tau=140~\mu$s was connected to each ring, and the calibration was verified by introducing a 10~mV voltage step across a 1~pF capacitance at the preamplifier input.  Prior to introducing O$_2$ dopant, we first verified proper detector operation and LEM gain under electron drift conditions in Ar (with 30\% CO$_2$ quench gas).  The $E=5.9$~keV gammas from the $^{55}$Fe source produce $n_0\equiv E/w=209$ electrons, based on $w=28.2$~eV. This value was obtained from a weighted average of the argon ($w=26.3$~eV) and CO$_2$ ($w=32.8$~eV) constituent w-values \cite{2008blum}, a result which is confirmed experimentally \cite{1963bortner}.  These drift as a swarm to the anode, are amplified by the LEM and in the absence of O$_2$ result in a single, fast output pulse from the preamplifier. Typical pulse rise times were measured to be $\mathcal{O}(10)$~ns.  From the known quantities of initial and measured charge, the total effective LEM gain was determined.  This is discussed further in Sec. \ref{sec:results}. A flow-through system with proportional-integral-derivative feedback pressure regulation was used to mitigate detrimental effects due to detector component outgassing.  Research-grade gas mixtures were used, and stable operation was observed on a timescale of several days.  All results presented in this article were obtained at a gas pressure $P=0.250\pm0.001$~bar and T~$=296\pm1$~K.

There are three distinct regions of electric field in the detector: (i) the $z_d=110$~mm drift region, with $E_d\sim0.05$~kV~cm$^{-1}$, (ii) the $z_A=0.432$~mm electron avalanche region $E_A$ in the LEM holes, with typical maximum electric fields of several tens of kV/cm \cite{2006shalem} and (iii) the $z_i=2$~mm signal induction region, with $E_i=1$~kV/cm.  $E_A$ is a function of radial displacement from the center of a hole. Values quoted in this work are given simply by $E_A=\Delta V / z_A$, where $\Delta V$ is the bias voltage across the LEM. The high electric field $E_A$ also provided the requisite energy to detach a captured electron from O$_2^-$.  The mechanism for this detachment process is thought to be collisional \cite{2010dion}.

\begin{figure*}[p]
\centering
\includegraphics[width=0.99\textwidth]{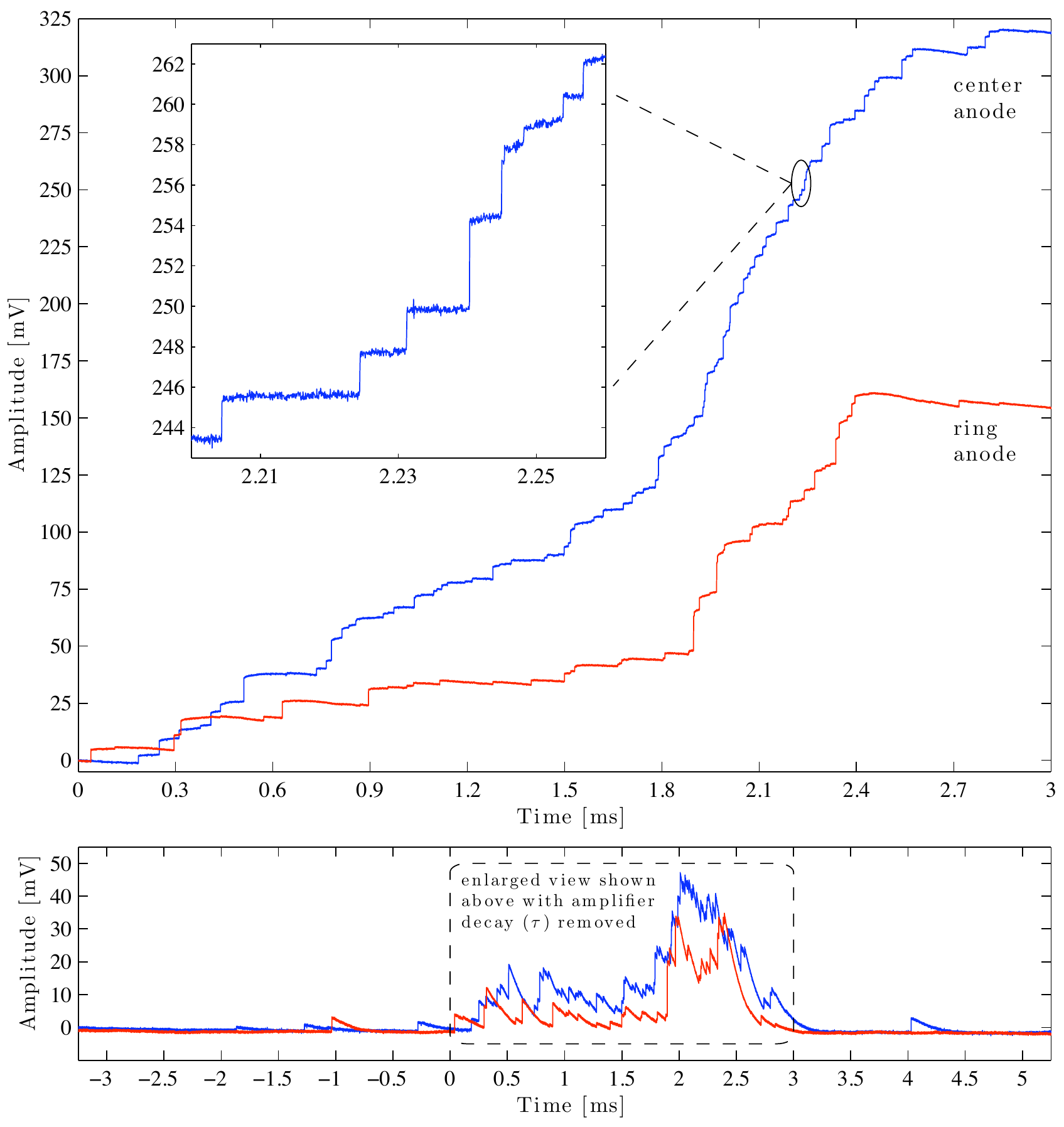}
\caption{Example event record from 5.9 keV energy deposition in Ar-CO$_2$ with 4\% O$_2$ dopant at $p=0.25$~bar.  The LEM was operated with $\Delta \mbox{V}=880$.  {\bf (Lower panel)} The event record as digitized. {\bf (Upper panel)} Enlarged view of the event record, after correcting for the charge preamplifier decay constant $\tau=140~\mu$s.  {\bf (Inset)} Additional detail over a 60~$\mu$s window. }
\label{fig2}
\end{figure*}

The amount of oxygen dopant required to fully capture all electrons created by 5.9 keV photons was first obtained from the measured reduction in pulse height, using electron drift operation.  We found that 1\% O$_2$ was sufficient to fully capture all electrons (though higher gas pressures required lower O$_2$ concentrations). However, the capture probability was sufficiently low that the resulting negative ion signal was distributed across about 10~ms.  We found that with 4\% O$_2$, the resulting signal was localized to within about 5~ms due to the more rapid capture of ionized electrons.  This gave an average 22~$\mu$s between subsequent ion arrival times at the LEM, sufficient to allow most ions to generate a distinct signal at the readout plane. In Sec. \ref{sec:cl} we quantify the fraction of missed ion pulses due to the proximity  in time of successive pulses. In order to ensure negligible event pileup, the $^{55}$Fe source rate was set at about 3~Hz, using a 0.88~mm diameter collimator.

\section{Data acquisition and analysis} \label{sec:results}

\subsection{Data acquisition} \label{sec:data}
Negative ion drift event records of 25~ms each were acquired at 20~MHz, with 12-bit vertical resolution.  A typical event obtained with an effective LEM gain of $3\times10^4$ ($\Delta V = 880$ across the LEM) is shown in Fig.~\ref{fig2}. Note that effective LEM gain refers to the total observed electron gain from a single electron, and accounts for any reduction due to secondary electrons which did not reach the anode.  The interesting portion of the event as digitized is shown in the lower panel.  An enlarged view with the effect of the charge preamplifier decay time removed is shown in the upper panel.  A total of $n_e=157$ charge pulse steps were counted: 103 on the center anode,  54 on the adjacent ring anode and 0 on the outer ring anode (not shown).  Each step is interpreted as corresponding to a single electron that was recovered from an O$_2^-$ ion, and amplified by the LEM. The possibility of pileup is discussed in Sec. \ref{sec:cl}.

\subsection{Analysis algorithm} \label{sec:algorithm}
Our analysis algorithm employs a simple edge-detection scheme: for a given sample $i$, the average value of the preceeding $n_b$ samples are subtracted from the average value of the following $n_a$ samples.  The analysis shown in this paper used $n_a=n_b=10$~samples.  As might be expected, increasing the number of samples contained in the average results in a higher signal to noise ratio, allowing smaller pulses to be discerned.  We found as many as 5\% more pulses with $n_a=n_b=50$~samples.  However, this comes at the expense of also increasing the number of pulses lost to pileup. In other words, the algorithm cannot resolve two single electron pulses that are separated in time by $<n_s/f_{ADC}=1~\mu$s. In this equation, $n_s=n_a+n_b$ is the number of samples used to find the edge of a single electron pulse and $f_{ADC}=20$~MHz. These pileup losses could be easily recovered by increasing $f_{ADC}$.

The algorithm is capable of finding arbitrarily small pulses in an event record. In order to make a meaningful determination of the number of recovered electrons, a pulse height threshold must be chosen. A simple choice is the pulse height value at which the slope of the distribution abruptly steepens, as indicated by the arrow in  Fig.~\ref{fig4} (inset). This point corresponds to the onset of spurious noise pulses, and is determined by the input capacitance to the preamplifier.  Accordingly, slightly higher thresholds were required on the larger-area anodes.  We note that Fig.~\ref{fig4} corresponds to the center anode. The thresholds employed in this analysis were [0.32~0.35~0.42]~mV, for the center, ring and outer ring anodes. We also investigated a threshold determination based on the observed pulse height distribution in a region where no signal is present, such as the initial $\sim1$ ms of an event record. Both methods lead to similar results.

Additionally, we explored a slightly more sophisticated analysis algorithm in which the existence of a pulse in the event record was determined by the 2nd derivative of the raw event record. In other words, a pulse was identified by a zero-crossing in the double differential. As with the simple scheme, the pulse candidate was still required to pass a threshold requirement. Similar results were obtained with both algorithms.

We note that the analysis avoids double-counting an electron pulse which may have had its signal split across two adjacent rings. This scenario was found to occur on approximately 5\% of pulses.

\subsection{Single-electron pulse height distribution} \label{sec:sephd}
\begin{figure}[ht]
\centering
\includegraphics[width=0.48\textwidth]{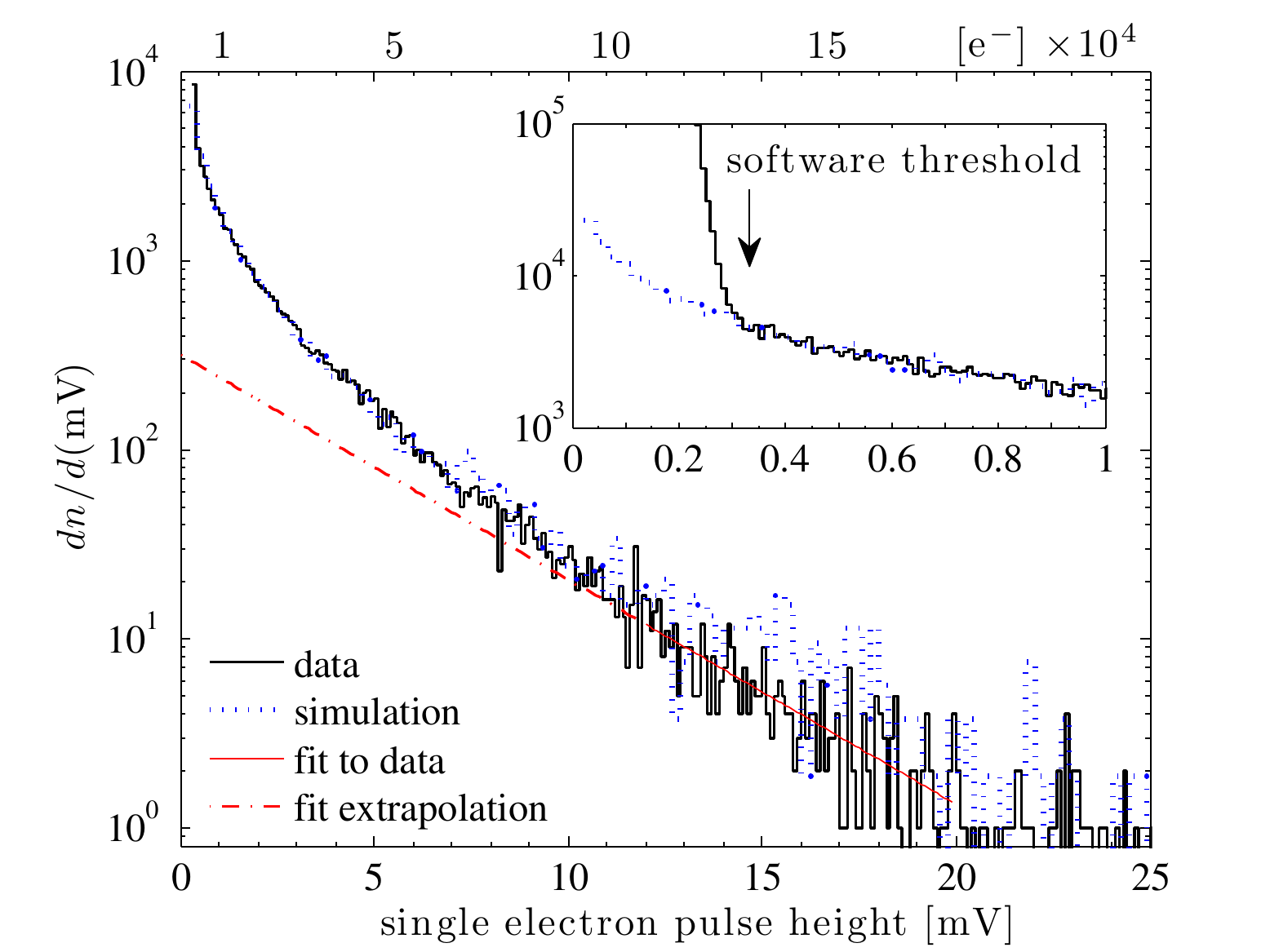}
\caption{The measured pulse height distribution of single electrons on the center anode, as exemplified in Fig.~\ref{fig2} (black steps). The function $y=\mbox{exp}(-x/g)$ was fit to the tail of the measured distribution (solid red curve, with small-$x$ extrapolation shown dashed).  The pulse height distribution obtained from the detachment model is also shown (dotted blue steps). Equivalent effective LEM gain is indicated along the top axis. {\bf (Inset)} The pulse height distribution near pulse detection threshold ($<1$~mV).}
\label{fig4}
\end{figure}

Figure \ref{fig4} shows the distribution of single electron pulse heights corresponding to the conditions exemplified in Fig.~\ref{fig2}.  The distribution of pulse heights resulting from avalanche multiplication is expected to be exponential \cite{1966cookson} and is often referred to as a Furry distribution.  We observe a steeper rise at small pulse heights.  The origin of this effect is a displacement of the $z$ coordinate at which an electron avalanche was initiated.

Consider the enlarged view of the LEM avalanche region shown in Fig.~\ref{fig1}.  In that example, the avalanche initiates at $z\approx-0.15$~mm.  In the absence of O$_2^-$ (electron drift),  the variation in the $z$ coordinate of avalanche initiation is $\sigma\approx0.05$~mm, as determined from our detachment model (described in the appendix). When O$_2^-$ is introduced, avalanches are (on average) initiated farther into the LEM, because electrons must first be detached from O$_2^-$ ions. This lowers the effective gain for a majority of electrons, as reflected in the reduced pulse height. This result is confirmed by our detachment model. A simulated pulse height spectrum obtained from the detachment model is also shown in Fig.~\ref{fig4}, normalized to the number of counts above the software threshold.

\subsection{Gain curves}
\begin{figure}[h]
\centering
\includegraphics[width=0.48\textwidth]{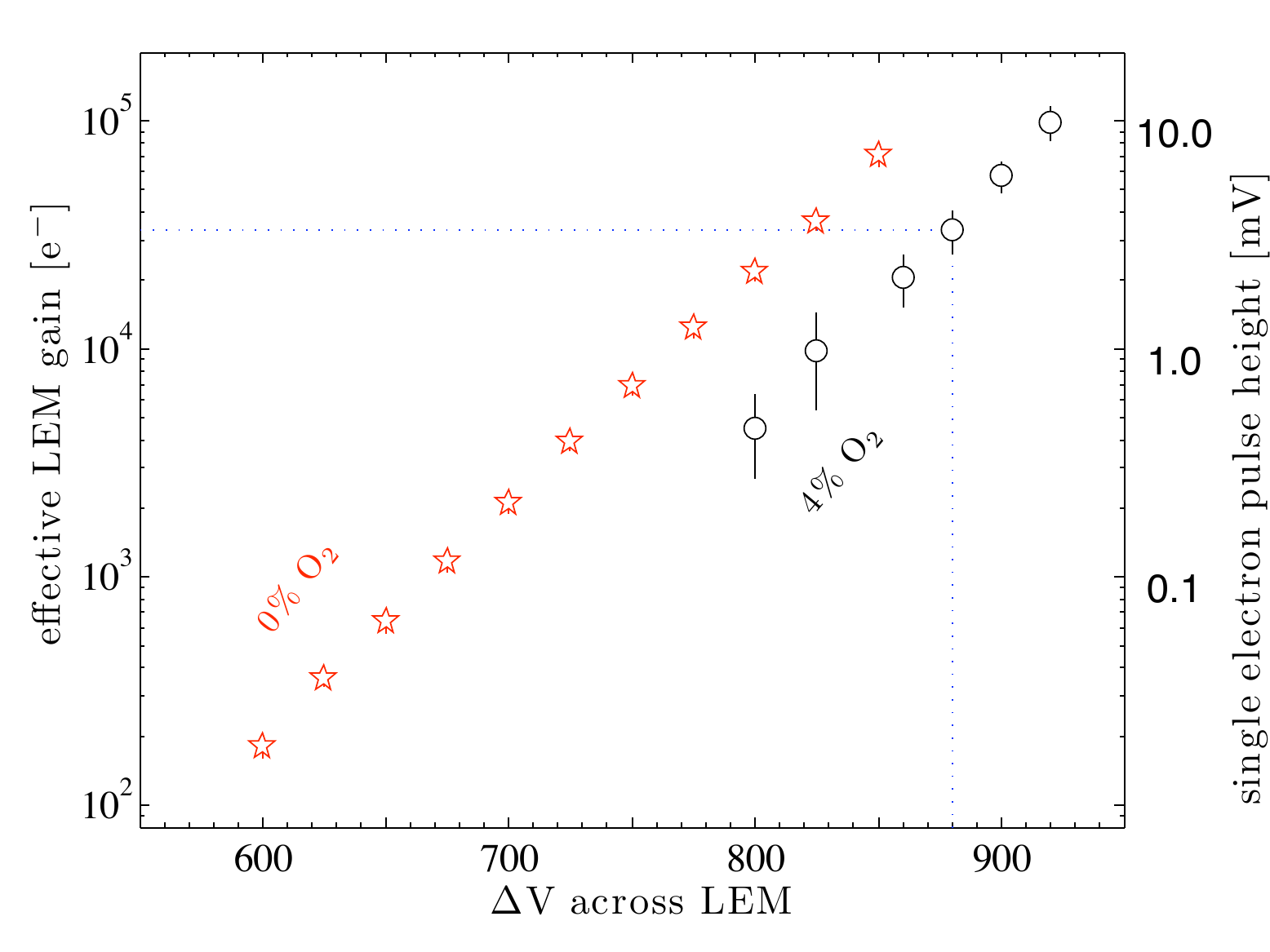}
\caption{Gain curves for the LEM.  The dotted crossed lines indicate the data point corresponding to the examples shown in Fig.~\ref{fig2}, Fig.~\ref{fig3} and Fig.~\ref{fig4}. Where not visible, uncertainty is smaller than the data point.}
\label{fig6}
\end{figure}

Gain curves for the LEM are shown in Fig.~\ref{fig6}, for 0\% (stars) and 4\% (circles) O$_2$.  The addition of O$_2$ lowers the effective gain at any particular voltage.  With 0\% O$_2$ the signal corresponding to the full-energy peak is a single pulse with a fast $\mathcal{O}$(10)~ns rise time. The gain curve for 0\% O$_2$ was obtained from that total pulse height, as described in Sec. \ref{sec:exp}.  The curve for 4\% O$_2$ was obtained from fitting $y=\mbox{exp}(-x/g)$ to the tail of the single electron pulse height distribution, as shown in Fig.~\ref{fig4}.  The decay constant $g$ of the exponential (in mV) was taken to indicate the mean gain at each voltage.  Considering Fig.~\ref{fig4}, it is expected that this procedure may result in a modest overstatement of the effective gain. We note that the present work does not require a precise knowledge of the gain; however, it is critical to verify that the gain is sufficiently large to see single electron pulses.

\section{Results}
\begin{figure}[h]
\centering
\includegraphics[width=0.48\textwidth]{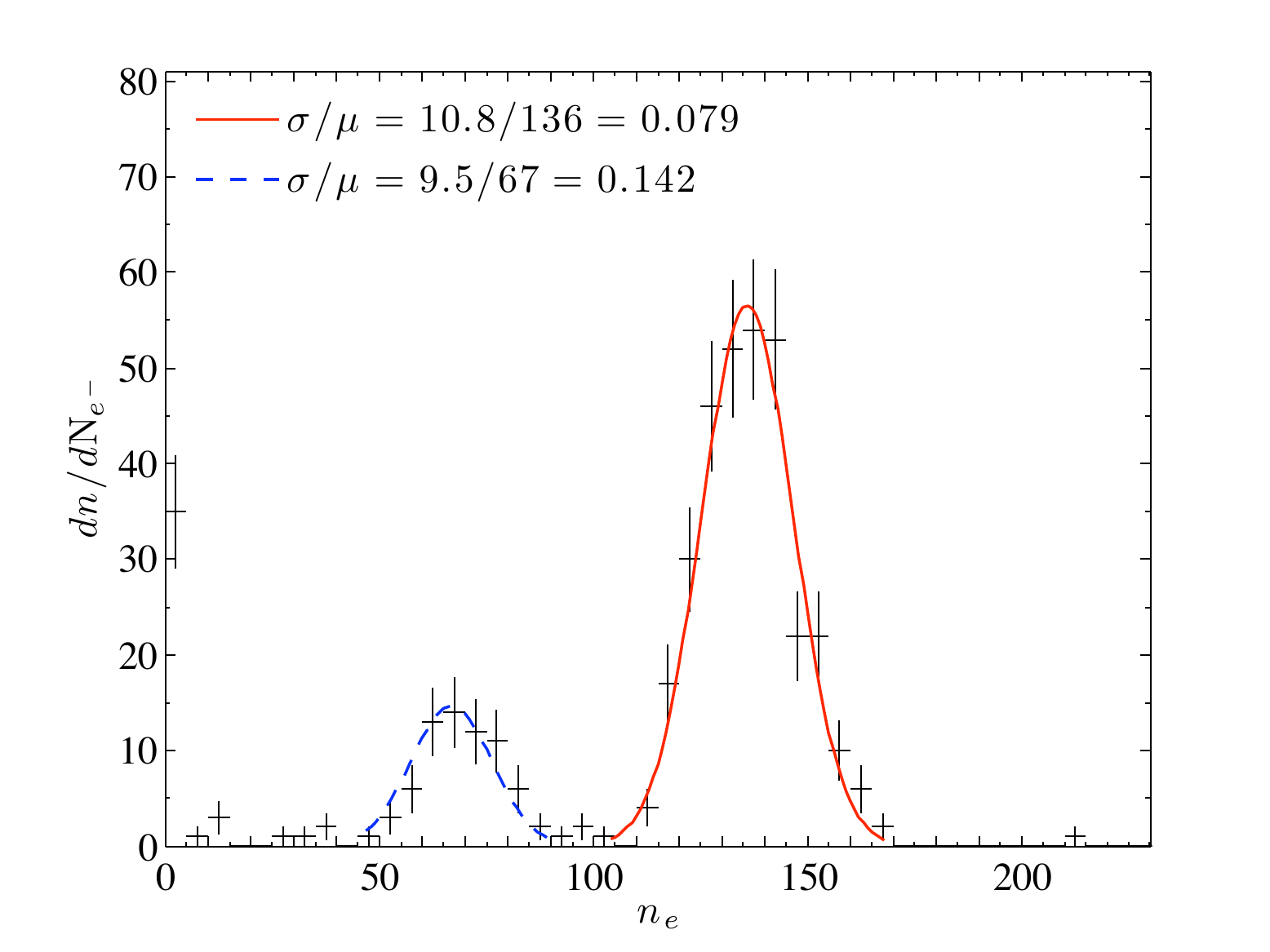}
\caption{The energy spectrum obtained from counting individual electrons recovered from O$_2^-$ ions, obtained in Ar-CO$_2$ at $p=0.25$~bar with 4\% O$_2$ dopant.  The LEM was operated with $\Delta \mbox{V}=880$. The full-energy peak is 5.9~keV and the argon k-shell escape peak is 3.1~keV.}
\label{fig3}
\end{figure}

A typical energy spectrum obtained from counting recovered single electrons on the center and ring anodes is shown in Fig.~\ref{fig3}. Software data selection cuts required $\leq1$ electron pulse in the first and last 3~ms of events, and $\leq5$ electron pulses on the outer ring anode. The energy resolution is discussed in Sec. \ref{sec:resolution}.

\subsection{Counting losses} \label{sec:cl}
\begin{figure}[h]
\centering
\includegraphics[width=0.48\textwidth]{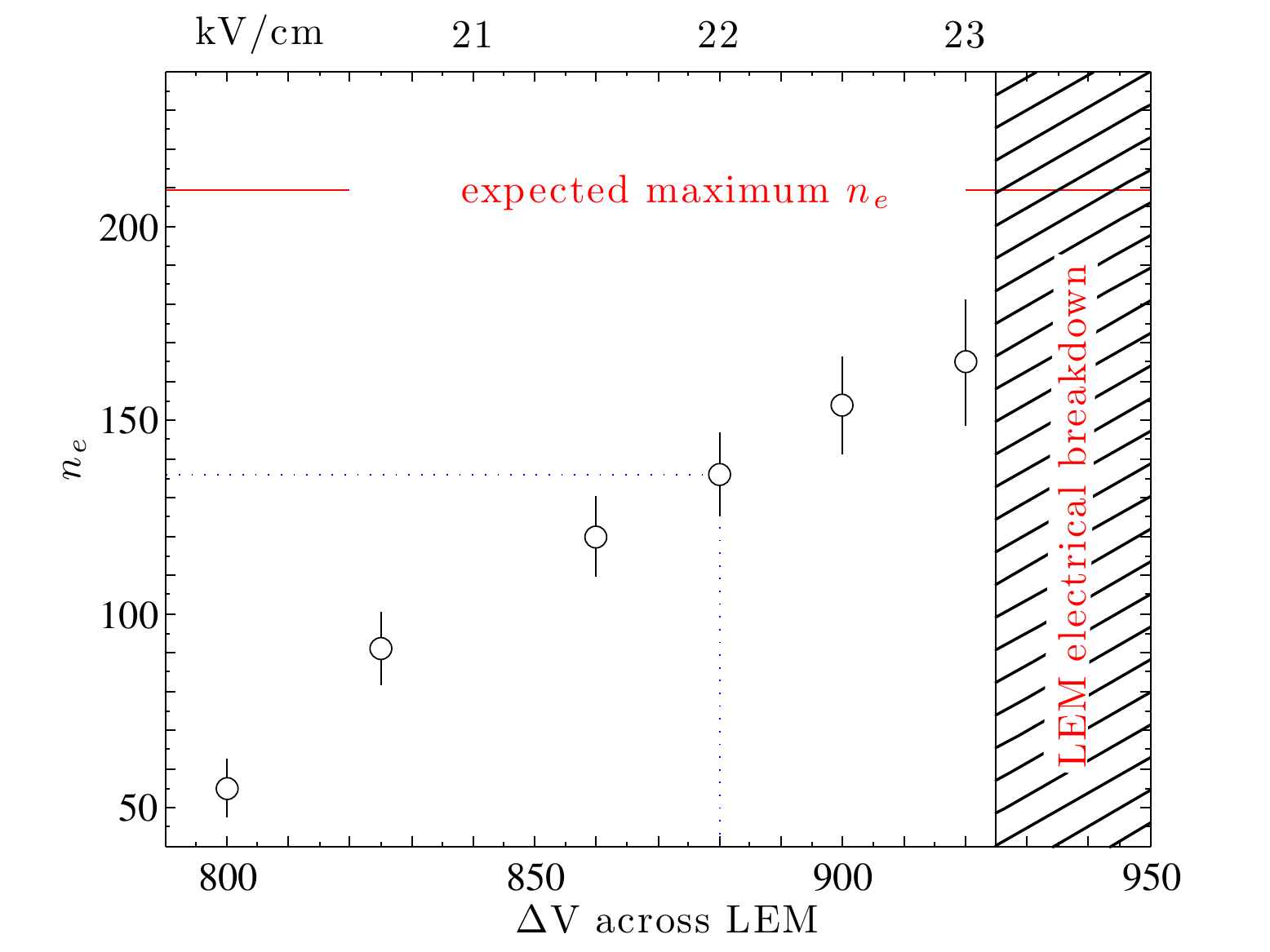}
\caption{Average number $n_e$ of counted electrons (with $1\sigma$ width) as a function of bias voltage $\Delta V$ across the LEM. Corresponding $E_A=\Delta V/z_a$ is indicated along the top.}
\label{fig7}
\end{figure}

Figure~\ref{fig7} shows the number $n_e$ of counted electrons as a function of LEM bias voltage.  A maximum electron counting efficiency of $0.78$ was obtained.  We consider several likely electron loss mechanisms:

\begin{enumerate}[(a)]
\item electrons which were not recovered from O$_2^-$.
\item electrons which were recovered from O$_2^-$ but did not produce an avalanche above threshold.
\item electrons which were recovered, produced an avalanche above threshold but were not resolved due to the finite value of $f_{ADC}$.
\item O$_2^-$ ions which collided with the top (in voltage) of the LEM, rather than being directed into one of the multiplication holes.
\item ``gas-phase chemistry'' \cite{2007nygren}. An example would be a dissociation reaction which left an appreciable fraction of e.g. O$^-$, whose higher electron affinity then precludes any hope of recovering the captured electron. Quantifying this effect is beyond the scope of this work, and is not discussed further.
\end{enumerate}

Addressing items (a) and (b), the trend of the data suggests that the full number of initial electrons could be recovered, if only $\Delta$V could be increased by another $\sim50$~V. However, in this bias regime frequent electrical breakdown makes the LEM gain structure unusable. 

Addressing item (b) in particular, we can use our detachment model to make a rough estimate of the fraction of single electron pulses which were lost below the software threshold. Looking at the simulated curve in Fig.~\ref{fig4} (inset), the predicted fraction of pulse heights below the software threshold is about 0.41. Under the experimental conditions shown there ($\Delta V=880$), the total fraction of electrons not counted was $1-136/209=0.35$. We see that the model slightly over-predicts the losses below software threshold. This is not surprising since there may be other factors (not addressed by the model) which influence the shape of the pulse height distribution. Nevertheless, the rough agreement is encouraging.

Addressing item (c), the fraction of pulses lost to pileup is approximately ${n_c}{n_s}/N$,  where $n_c$ is the number of pulses occurring on a channel during the $N$ samples spanning a typical event. As shown in Fig.~\ref{fig2}, typically $N\approx6\times10^4$~samples. From Sec. \ref{sec:algorithm}, $n_s=20$. We must have $n_c \leq n_e$, where $n_e$ is the total number of pulses in an event. The limiting case occurs if all the electrons arrive at the same channel. Taking $n_c=136$ as in Fig.~\ref{fig3}, this simple estimate implies that the typical maximum fraction of electrons lost to pileup is 0.04. The result is confirmed by a detailed analysis of the time between successive pulses in all events in our data sample.

Addressing item (d), the next obvious parameter which might be increased is the induction field strength $E_i$. Previous studies \cite{2006shalem} have shown that the transparency of a LEM for electrons depends on $E_i$. For LEM bias values $\Delta V\lesssim 880$, we found that by doubling $E_i$ from 1~kV/cm to 2~kV/cm we obtained a 20\% increase in the number of recovered electrons. Further increase of $E_i$ did not yield additional recovered electrons. Interestingly, when operating at higher bias values $\Delta V> 880$ with $E_i=2$~kV/cm, the maximum number of recovered electrons was still limited to about $\times0.78$. 

An additional complication of operation at $E_i>1$~kV/cm is a distortion of the pulse shape. With $E_i=1$~kV/cm individual pulse shapes were as expected: a voltage step with fast $\mathcal{O}(10)$~ns rise time, followed by recovery with $\tau=140~\mu$s. With $E_i=2$~kV/cm, we observed the same initial fast rise time, followed immediately by a slow $\sim3~\mu$s rise time, followed by recovery with the expected time constant. The step height of the slow rise time was less than half the step height of the fast rise time, and was observed even in the absence of oxygen. This leads to the interpretation that it is due to positive ion movement between the lower (in voltage) electrode of the LEM and the anode. Physically, this implies that the electron avalanche continued for several hundred microns beyond the lower electrode plane of the LEM.

We have assumed that $w$-value is unaffected by the presence of the oxygen dopant. This seems reasonable since oxygen serves to remove, rather than add, ionized electrons. However, for completeness we note that if the $w$-value were higher (lower) by \emph{e.g.} 2~eV, the predicted resolution shown in Fig. \ref{fig8} would be lower (higher) by only a few percent. This is a small effect $-$ roughly equivalent to the line width of the curve. At the same time, the expected maximum $n_e$ shown in Fig.~\ref{fig7} would decrease (increase) by about 7\%.

\subsection{Energy resolution} \label{sec:resolution}
\begin{figure}[h]
\centering
\includegraphics[width=0.48\textwidth]{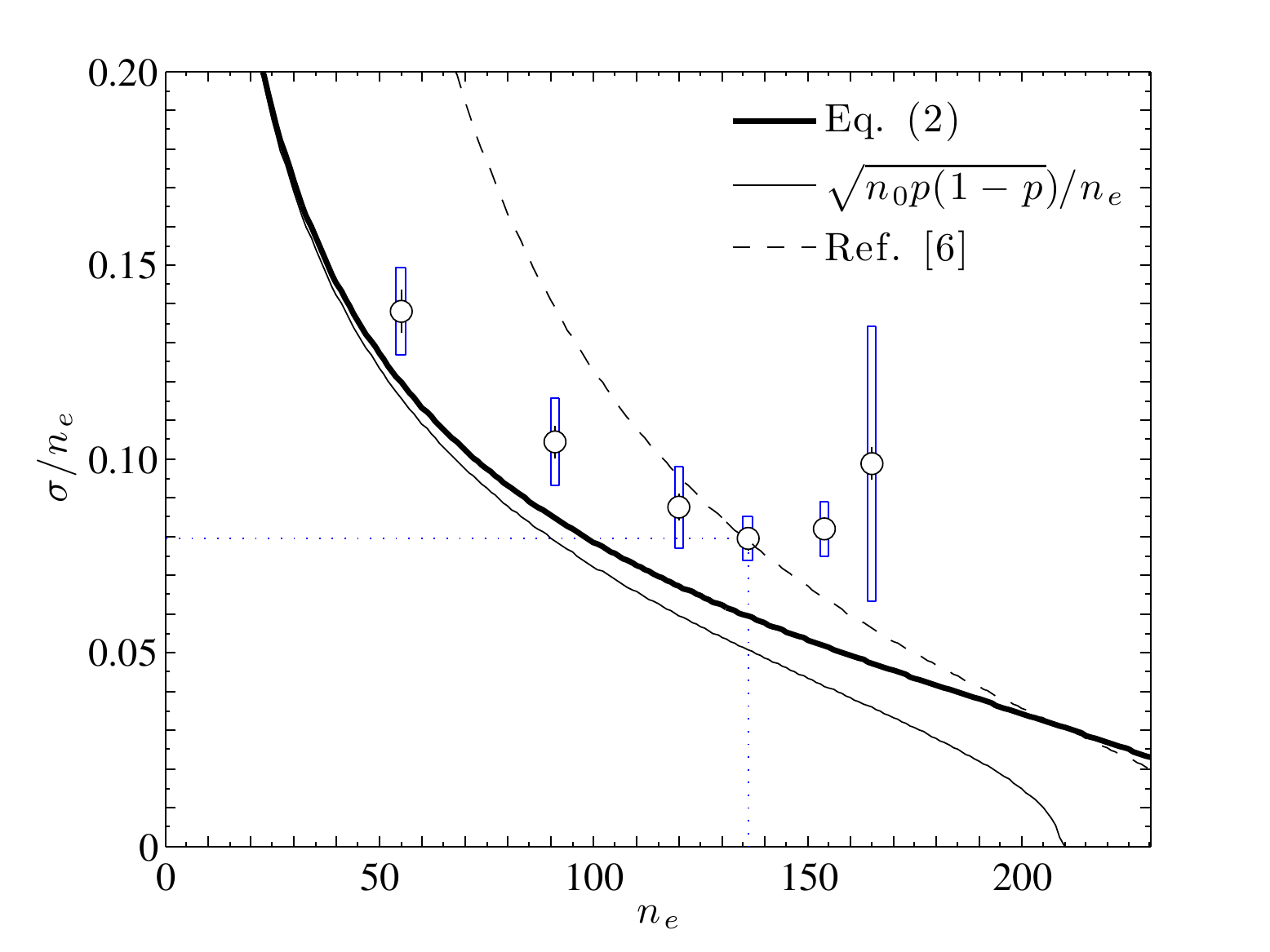}
\caption{Observed energy resolution (circles) compared with the expectation for a binomial process (thin solid), the prediction from Ref. \cite{2007nygren} and our prediction as given by Eq. \ref{eq:res} (thick solid). Statistical uncertainty is indicated by the thin vertical bars, and systematic uncertainty (discussed in the text) is indicated by the vertical rectangles.}
\label{fig8}
\end{figure}

The best energy resolution we achieved under electron drift conditions (in the absence of O$_2$) was  $\sigma/\mu=0.11$. In Fig.~\ref{fig8} we show the energy resolution $\sigma/n_e$ as a function of $n_e$, as obtained from negative ion drift. Statistical uncertainty is indicated by the thin vertical bar, and in several cases is smaller than the data point. Sources of systematic uncertainty, indicated by the larger vertical rectangle, include (1) varying $n_{a,b}$ between 10 and 20 samples, (2) variation due to binning and fitting and (3) differences in results between two independent analyses. The last item implicitly accounts for variation in the determination of the pulse detection software threshold.

The expected detector resolution for electron counting is
\begin{equation} \label{eq:res}
\sigma/n_e = \sqrt{\frac{\mbox{F}}{n_0} + \frac{(1-p)}{n_0p}}
\end{equation}
where $F=0.20$ \cite{1984kase} is the Fano factor in gaseous argon. This equation states that the resolution depends on the Fano fluctuation in initial number $n_0$ of ionized electrons, and on the binomial process by which electrons are recovered from O$_2^-$ and amplified. As mentioned previously, we define $n_e=n_0p$, where $p$ is the total probability for an electron to be counted. 

The agreement between Eq. \ref{eq:res} and the data is not satisfactory for any value of $n_e$. We are therefore lead to consider other processes which remain unaccounted in our analysis. We note that if the measured $n_e$ values were $\sim30\%$ smaller, our data would appear to agree well with Eq. \ref{eq:res}. This leads to the suspicion of a spurious double-counting of recovered electrons.   We have already removed the possibility of double-counting pulses across adjacent anodes. There remains the possibility of double-counting due to some sort of feedback mechanism. For example, we know that copious UV photons are generated during the electron avalanche. Could there be a small probability that one of these photons eject an electron from {\it e.g.} the wall of the LEM, leading to a spurious electron pulse? We have investigated this possibility and found no concrete evidence to support the hypothesis, but this does not allow us to rule it out. For completeness we note that the discrepancy we observe cannot be accounted for by any reasonable variation in the Fano factor. We also note that our data do not support the theoretical energy resolution proposed in \cite{2007nygren}.

\section{Summary and future work}
We have demonstrated that improved energy resolution can be achieved in a negative ion time projection chamber by individually counting the electrons produced by ionizing radiation. The initial results are encouraging, but many questions remain to be addressed. Higher electron gain and a higher electric field $E_A$ are clear priorities. Future work will study the electron counting performance using proportional wire and MICROMEGAS gain structures, as well as optimal gas mixtures (including our desired target isotope, xenon) and operation at increased gas pressure.

\section*{Appendix: a simplified electron detachment model} \label{sec:detmod}
In order to further understand the physics underlying the single electron pulse height distribution discussed in Sec. \ref{sec:sephd}, we employed a simplified model of the detachment process.  It was assumed that  O$_2^{-}$ ions drift in an ideal gas of neutral argon molecules in thermal equilibrium. The ions make elastic collisions described by a cross section calculated assuming a long-range polarization interaction with the neutral atom \cite{1958gioumosis}. They are accelerated by a constant electric field between collisions.  The electron detaches from the ion when the center of mass energy of the ion-neutral system (between collisions) exceeds a threshold $\varepsilon_D$. Using momentum transfer theory \cite{1988mason}, the model produced a lifetime for electron detachment for a given electric field strength and $\varepsilon_D$, and a distribution of detachment probabilities for a given electric field profile. It would seem natural to associate $\varepsilon_D$ with the electron affinity, which is $\varepsilon=0.45$~eV \cite{2003ervin} for O$_2$.  Based on experimental determinations of the collision cross section for detachment of both O$_2^{-}$ ions \cite{1970wynn} and several types of halogen ions \cite{1978smith}, it appears that $\varepsilon_D$ can often exceed the electron affinity by $\times2$ or more.  

\begin{figure}[ht]
\centering
\includegraphics[width=0.48\textwidth]{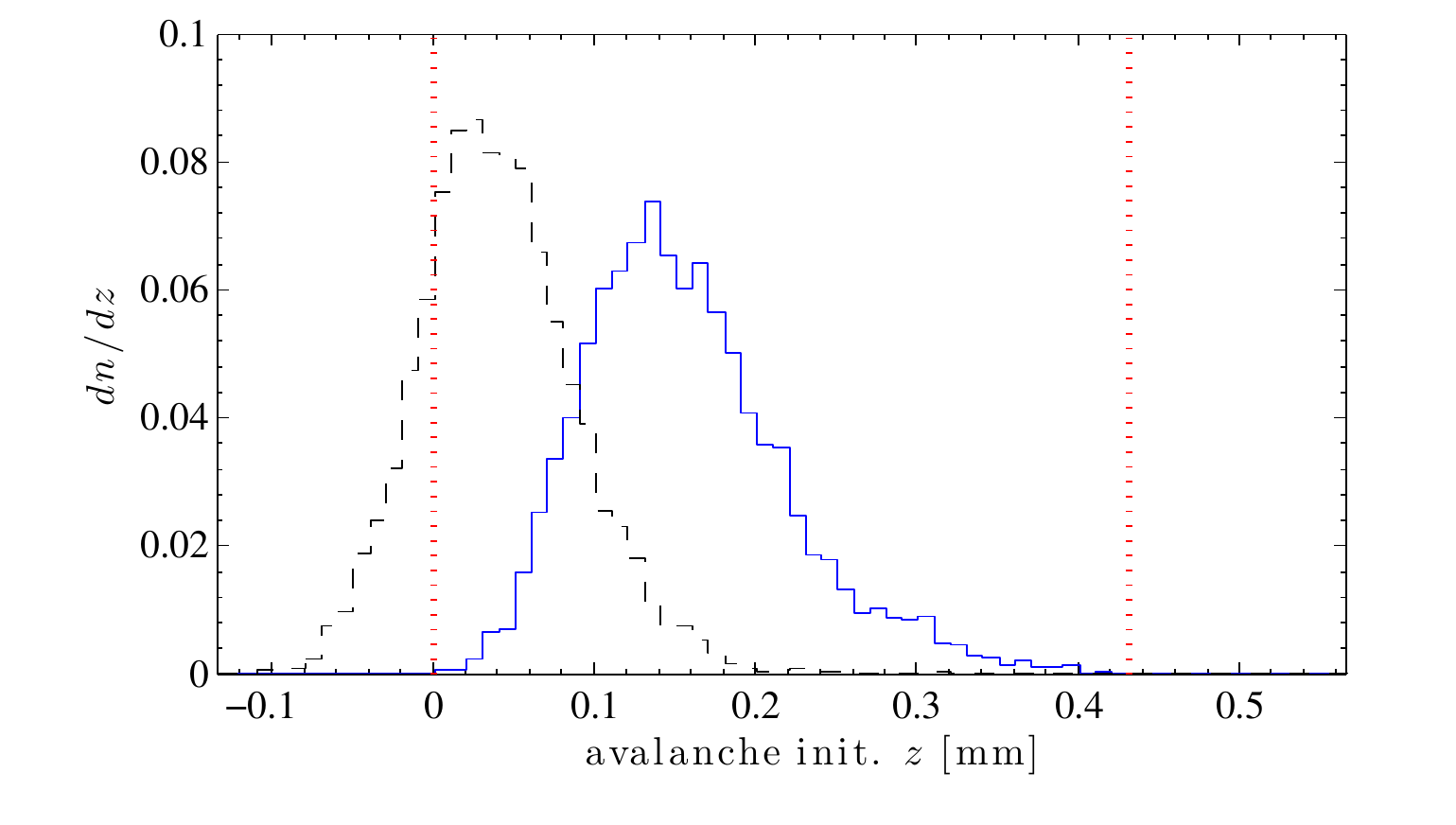}
\caption{Simulated  distribution of $z$ coordinate at which electron avalanches begin, with (solid) and without (dashed) O$_2$. The vertical dotted lines indicate the physical extent of the LEM.}
\label{fig5}
\end{figure}

We therefore treated $\varepsilon_D$ as a free parameter. Different values of $\varepsilon_D$ resulted in different $z$ detachment profiles. These were incorporated into Monte Carlo simulation to study the effects of ion detachment on single-electron pulse height distributions. 
 The simulated pulse height distribution for the avalanche gain for $\varepsilon_D=3$~eV is shown in Fig.~\ref{fig4}. The point of avalanche initiation which resulted from the detachment profile (also obtained with $\varepsilon_D=3$~eV) is shown in Fig.~\ref{fig5}. The results were obtained for a gas mixture of ArCO$_2$, using Garfield$++$ \cite{2011schindler} in conjunction with Gmsh \cite{gmsh} and Elmer \cite{elmer}.  The simulated distribution for electron drift is also shown.  We note that by default, the predicted gain was about $\times10$ too low. This may be partially due to Penning processes which are not accounted for in the Garfield$++$ simulation \cite{veenhof}.  As an interim solution we explored the effect of varying the default Townsend coefficient data input to Garfield$++$, and found reasonable agreement (as shown here) with a uniform increase of 19\%.  This is well within the range allowed by experimental uncertainty \cite{2003auriemma,1993sharma}. 

\section*{Acknowledgments}
This work was performed under the auspices of the U.S. Department of Energy by Lawrence Livermore National Laboratory in part under Contract DE-AC52-07NA27344. J. Renner also acknowledges support from the Stewardship Science Graduate Fellowship (grant number DE-FC52-08NA28752). We are indebted to A. Goldschmidt, D. Nygen, J. Siegrist and D. Snowden-Ifft for useful discussions. 


\end{document}